\documentclass[11pt]{emulateapj}\usepackage{times}

\usepackage[usenames]{color}

\usepackage{gensymb}
\usepackage{graphics}
\usepackage{graphicx}
\usepackage{epstopdf}
\usepackage{url}
\usepackage{hyperref}
\usepackage{breakurl} 
\usepackage[stable]{footmisc}
\usepackage{lineno}

\newcommand{\be}{\begin{equation}}
\newcommand{\ee}{\end{equation}}
\newcommand{\ba}{\begin{eqnarray}}
\newcommand{\ea}{\end{eqnarray}}
\newcommand{\bc}{\begin{center}}
\newcommand{\ec}{\end{center}}

\newcommand{\psr}{PSR B1259--63}

\begin{document}

\slugcomment{ApJ, in press}

\title{\textsc{Gamma-ray flare activity from \psr\ during 2014 periastron passage and comparison to its 2010 passage}}
\author{G. A. Caliandro\altaffilmark{1,2}, C. C. Cheung\altaffilmark{3}, J. Li\altaffilmark{4}, J. D. Scargle\altaffilmark{5}, D. F. Torres\altaffilmark{4,6}, K. S. Wood\altaffilmark{3}, M. Chernyakova\altaffilmark{7,8}}
\altaffiltext{1}{W. W. Hansen Experimental Physics Laboratory, Kavli Institute for Particle Astrophysics and Cosmology, Department of Physics and SLAC National Accelerator Laboratory, Stanford University, USA; caliandr@slac.stanford.edu }
\altaffiltext{2}{Consorzio Interuniversitario per la Fisica Spaziale (CIFS), Italy}
\altaffiltext{3}{Space Science Division, Naval Research Laboratory, Washington, DC 20375, USA; kent.wood@nrl.navy.mil}
\altaffiltext{4}{Institute of Space Sciences (CSIC-IEEC), Campus UAB, c. de Can Magrans s/n, 08193 Bellaterra, Barcelona, Spain; jian@ieec.uab.es}
\altaffiltext{5}{Space Sciences Division, NASA Ames Research Center, Moffett Field, CA 94035-1000, USA}
\altaffiltext{6}{Instituci\'o Catalana de Recerca i Estudis Avan\c{c}ats (ICREA),  08010 Barcelona, Spain}
\altaffiltext{7}{Dublin City University, Dublin 9, Ireland}
\altaffiltext{8}{School of Cosmic Physics, Dublin Institute for Advanced Studies, Dublin 2, Ireland}

\begin{abstract}

PSR B1259--63/LS 2883 is a gamma-ray binary system containing a radio pulsar in a highly elliptical $\sim$ 3.4-year orbit around a Be star. In its 2010 periastron passage,
multiwavelength emission from radio to TeV was observed, as well as an unexpected GeV flare measured by the \emph{Fermi} Large Area Telescope (LAT). Here, we report the results of LAT
monitoring of PSR B1259--63 during its most recent 2014 periastron passage. We compare the gamma-ray behavior in this periastron with the former in 2010 and find that \psr\ shows a recurrent GeV flare. The similarities and differences in the phenomenology of both periastron passages are discussed.

\end{abstract}


\section{Introduction}

Gamma-ray binaries constitute a small subgroup of X-ray binaries hosting O/B stellar companions (Dubus 2013). They emit modulated radiation at the orbital period.
In fact, the gamma-ray emission from these systems dominates their spectral energy distributions.
Among all members in this group, \psr\ is the only one for which the compact object's nature is known:
a 47.76 ms radio pulsar orbiting a Be star (LS 2883) with a period of $\sim 1236.7$ days. It is located at a distance of 2.3 kpc (Johnston et al. 1992, 1994; Negueruela et al. 2011; Shannon et al. 2014).
Because of the high inclination angle ($i_{D}$) between the pulsar orbital plane and the Be star circumstellar disk ($i_{D}\sim 10^{\circ}-40^{\circ} $, Melatos et al. 1995), the circumstellar disk is crossed twice when the pulsar moves in one full highly elliptical (e~$\sim$~0.87) orbit. During these crossings, multiwavelength emission results from the interaction between the relativistic pulsar wind and the stellar wind of the companion (Chernyakova et al. 2006, 2009, 2014; Aharonian et al. 2009; Abdo et al. 2011).

Around the  2010 periastron passage of \psr\ (2010 December 14), enhanced gamma-ray emission as well as a month-long GeV flare were observed (Abdo et al. 2011; Tam et al. 2011).
During the GeV flare, \psr\ was characterized by an extremely high conversion efficiency of the pulsar spin-down power into gamma-ray emission.
The physical origin of the flare is still under debate (e.g., Khangulyan et al. 2012; Kong et al. 2012; Dubus \& Cerutti 2013; Chernyakova et al. 2014).
After its 3.4-year orbit period, \psr\ approached periastron again (2014 May 4). We now report the observation of a flare following the 2014 periastron at a time compatible with the extrapolated expectation following the 2010 periastron. The initiation of the flare was reported in Astronomer's Telegrams\footnote{\url{http://www.astronomerstelegram.org/}} by Malyshev et al. (2014); Tam et al. (2014); Wood et al. (2014). Detailed results of our monitoring campaign in GeV and X-rays, analyzing the \emph{Fermi} Large Area Telescope (LAT; Atwood et al. 2009) data and \emph{Swift}/XRT data, respectively, are given. We also provide a comparison with a reanalysis of the 2010 periastron data using the latest LAT instrument response functions (IRF) and diffuse models.\footnote{In the process of preparing this paper, a similar study by Tam et al. (2015) was presented. While their main results are consistent with ours, we have included a more detailed comparison between 2010 and 2014 flares as well as pulsar timing analysis.}

\section{Observations and data analysis}

The analysis of \emph{Fermi}-LAT data was performed using the
\emph{Fermi} Science Tools\footnote{\url{http://fermi.gsfc.nasa.gov/ssc/}}, 09-34-01 release.
For the 2010 and 2014 periastron passages the analysis was carried out with Pass 7 reprocessed data belonging to the SOURCE event class\footnote{\url{http://fermi.gsfc.nasa.gov/ssc/data/analysis/documentation/Pass7REP_usage.html}}.

\begin{figure}[t]
\centering
\includegraphics[angle=0, scale=0.45] {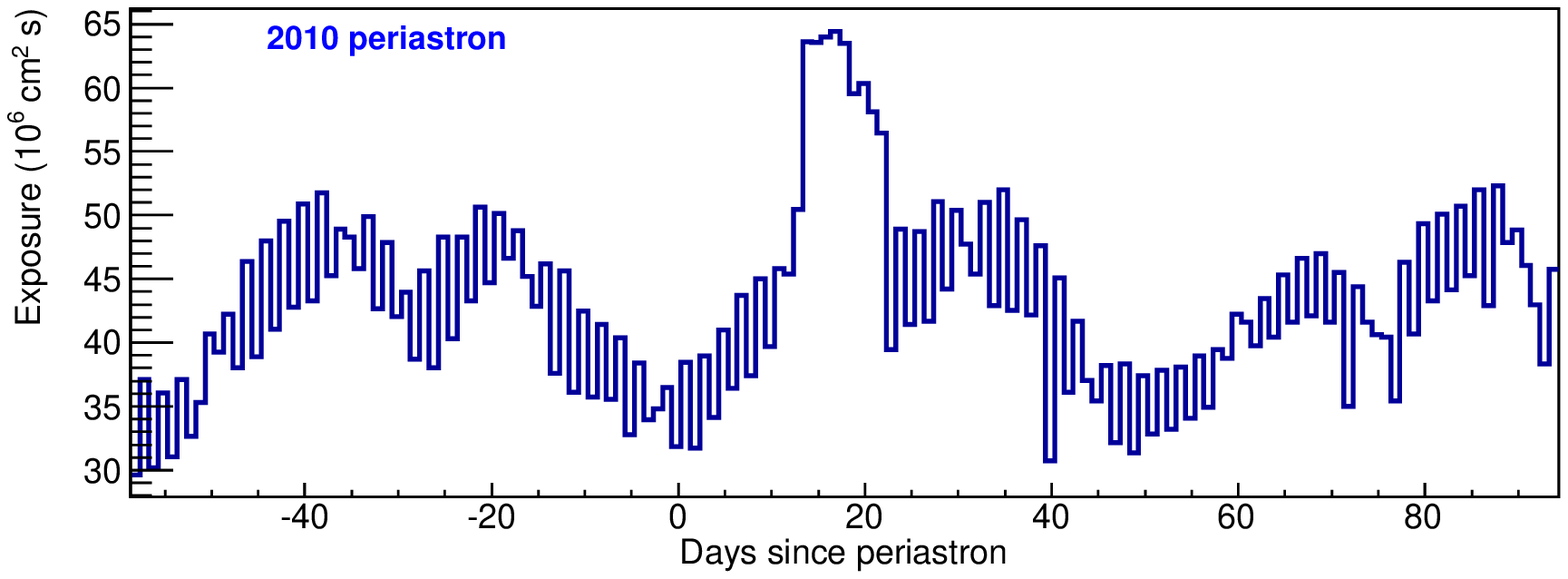}
\includegraphics[angle=0, scale=0.45] {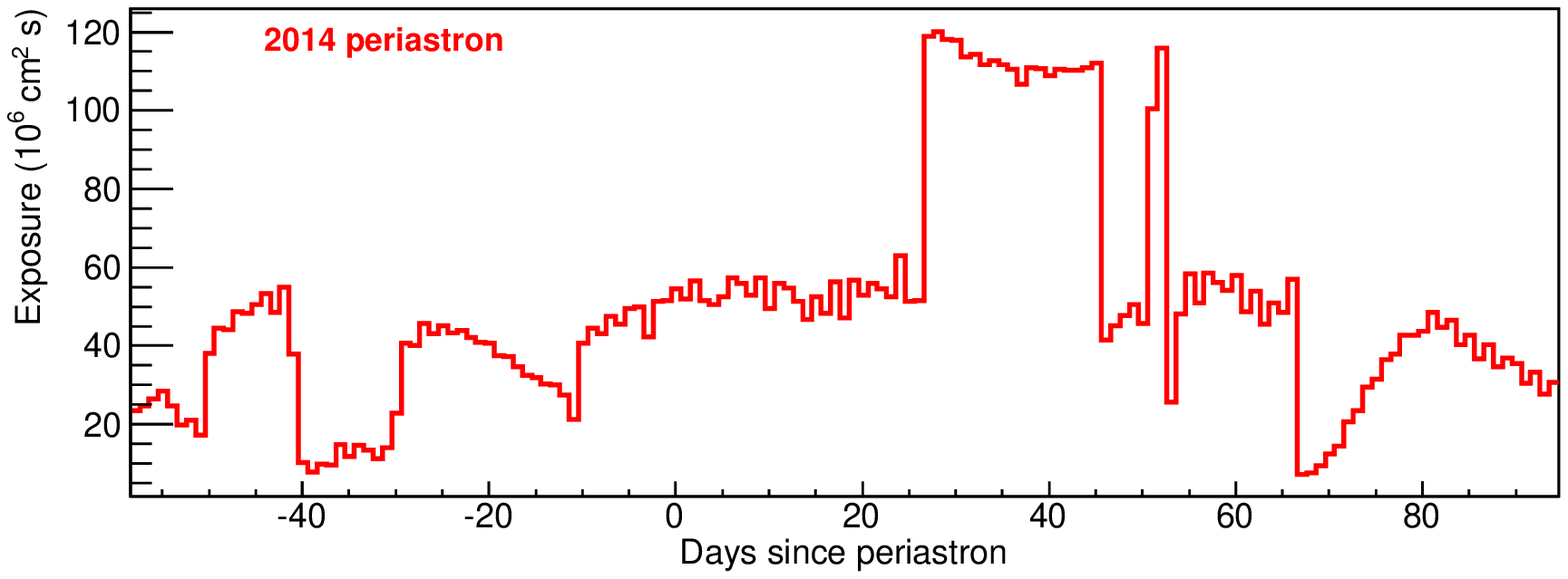}
\caption{\emph{Fermi}-LAT exposure of PSR B1259--63 during the 2010 (panel a) and 2014 (panel b) periastron passages.}
\label{expo}
\end{figure}

All gamma-ray photons within an energy range of 0.1--100 GeV and within a circular region of interest (ROI) of
10\degree\ radius centered on \psr\ were used for this analysis. To reject contaminating gamma rays from the
Earth's limb, we selected events with zenith angle $<$ 100\degree\/.
\emph{Fermi}-LAT had triggered Target of Opportunity (ToO)
observations in 2014 starting 27 days after the periastron, and lasting 19
days. During the ToO, the exposure of PSR B1259--63 was increased by a factor
of $\sim$2 compared to normal observations (Figure \ref{expo}, panel b).
In contrast, during the 2010 periastron passage a modified sky survey had
commenced 13 days after periastron. In this mode, which lasted 10 days, the
southern hemisphere receives 30\% extra exposure (Figure \ref{expo}, panel a).
The gamma-ray flux and spectral results of \psr\ presented in this work were calculated by performing a binned
maximum likelihood fit using the Science Tool \emph{gtlike}.
The spectral-spatial model constructed to
perform the likelihood analysis includes Galactic and isotropic diffuse emission components as well as known gamma-ray sources
within 15\degree\ of \psr\ based on a catalog internal to the collaboration (now released as the 3FGL catalog; Acero et al. 2015).
The spectral parameters were fixed to the catalog values, except
for the sources within 3\degree\ of \psr\/\footnote{\burl{http://fermi.gsfc.nasa.gov/ssc/data/analysis/scitools/source_models.html}}. For these latter sources, the flux normalization
was left free. We verified that changing 3\degree\ to 5\degree\ produced consistent results.
\psr\ itself was modeled as a single power-law with all spectral parameters allowed to vary.
The contribution of the Galactic and isotropic diffuse emissions within the analyzed ROI was estimated performing a preliminary maximum likelihood fit. For the 2010 periastron event, this fit included data from 5 months prior to 1 year after periastron.  For 2014, this fit included data from 5 months prior to periastron. The resulting scale factors of the isotropic and Galactic diffuse templates were kept fixed in the analysis that follows.
The 2010 and 2014 periastron times used in this paper are MJD 55544.693781 (2010 December 14 16:39:02.678) and MJD 56781.418307 (2014 May 4 10:02:21.725), which were derived from the most recent measurement of the binary system's orbital parameters (Shannon et al. 2014).

\emph{Swift}/XRT also monitored \psr\ during the 2014 periastron passage in Photon Counting (PC) and Window Timing (WT) modes\footnote{\url{https://www.swift.psu.edu/xrt/software.html}}.
We selected PC data with event grades 0--12 and WT data with event grades 0--2 (Burrows et al. 2005). Source events were accumulated within
a circular region with a radius of 30 pixels (1 pixel = 2.36 arcsec). Background events were accumulated within a circular, source-free region with a radius of 60 pixels.
Exposure maps were generated with the task XRTEXPOMAP. Ancillary response files were generated with XRTMKARF, which accounts
for different extraction regions, vignetting, and point spread function corrections. We analyzed the \emph{Swift}/XRT 0.1--10 keV data by HEAsoft version 6.14, \footnote{\url{http://heasarc.nasa.gov/lheasoft/}} and spectral fitting was performed using XSPEC V.12.8.1.

\begin{figure*}[t]
\centering
\includegraphics[angle=0, scale=0.45] {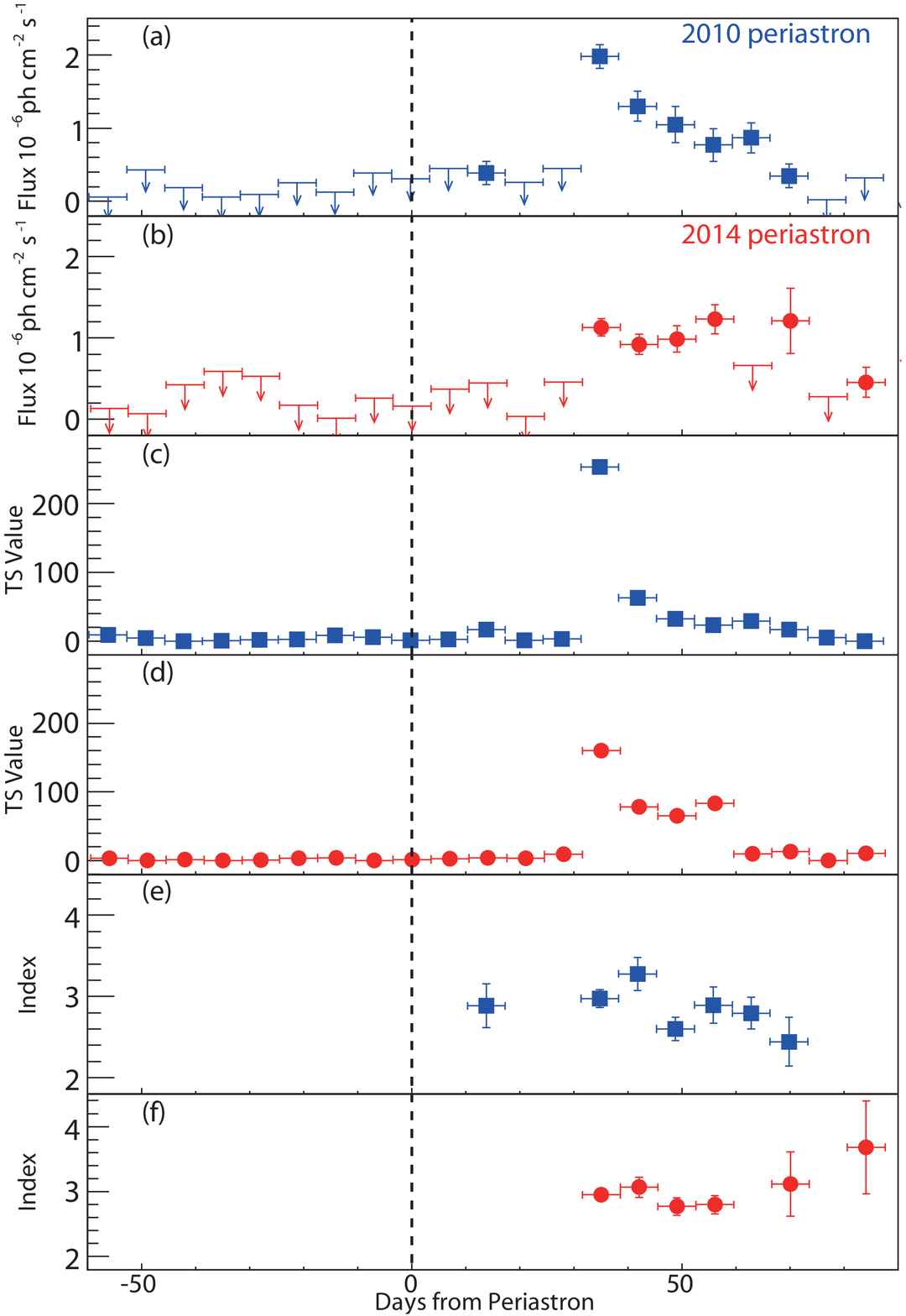}
\includegraphics[angle=0, scale=0.45] {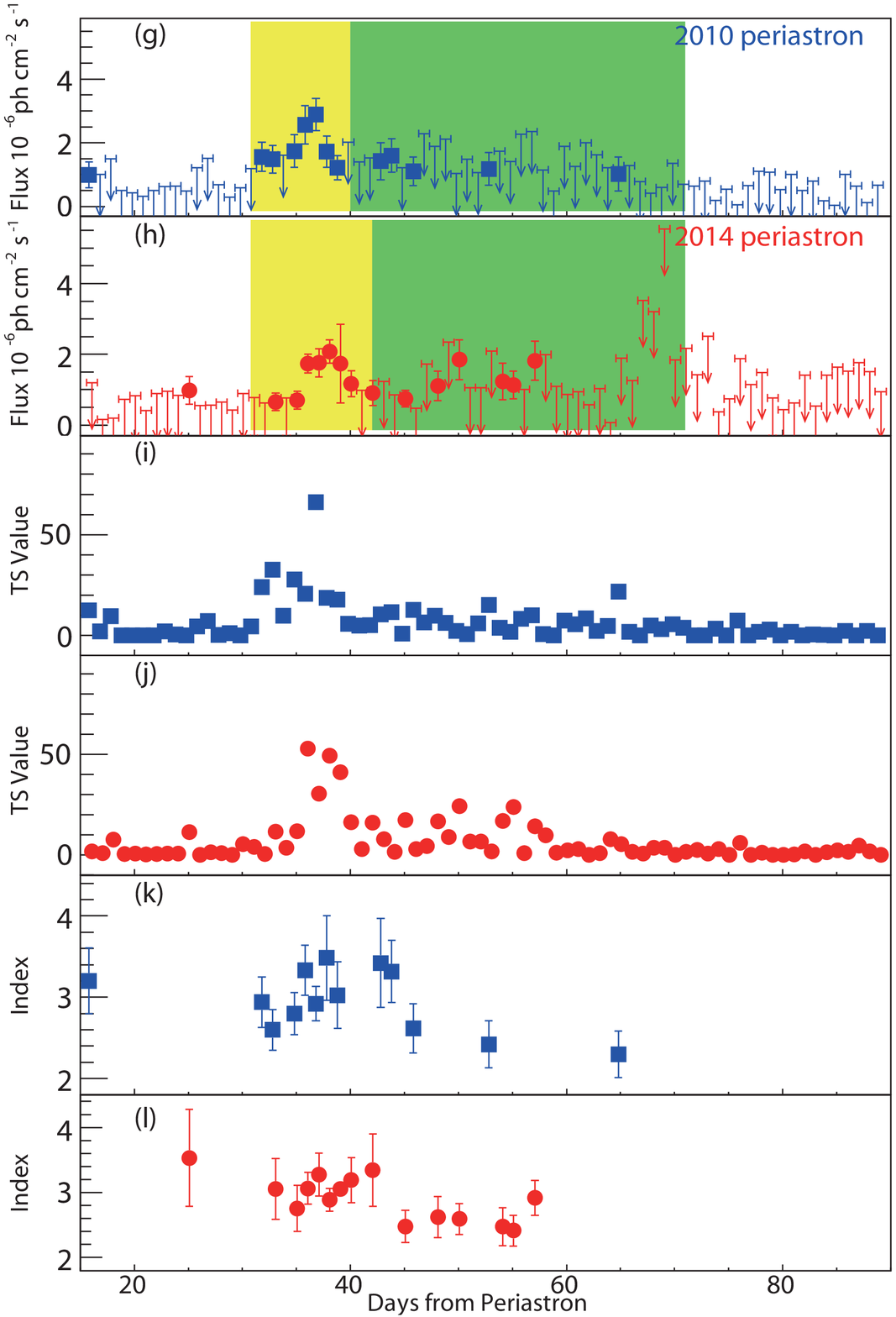}
\caption{Weekly (left panels) and daily (right panels) gamma-ray flux (a \& b, g \& h), TS value (c \& d, i \& j), and photon index (e \& f, k \& l) of PSR B1259--63 during the 2010 (blue) and 2014 (red) periastron passages. The dashed black line in left panels is a fiducial zero indicating the time of periastron. The shaded yellow and green regions indicated in panels  g\& h are the defined Peak and Tail periods.}
\label{mos-1}
\end{figure*}

\section{Light curve}

\subsection {GeV flares}

During \psr\/'s 2010 periastron passage, bright GeV emission was observed $\sim$~30--80 days after periastron (Abdo et al. 2011).
In the 2014 periastron passage, we monitored the gamma-ray emission from \psr\ on daily and weekly timescales starting 60 days before periastron. For the time periods when \psr\ was not detected with the LAT (Test Statistic (TS)\footnote{The Test Statistic is defined as TS=$-2 \ln (L_{max, 0}/L_{max, 1}) $, where $L_{max, 0}$ is the maximum likelihood value for a model without an additional source (the ``null hypothesis") and $L_{max, 1}$ is the maximum likelihood value for a model with the additional source at a specified location. A larger TS indicates that the null hypothesis is incorrect (i.e., a source really is present). As a basic rule of thumb, the square root of the TS is approximately equal to the detection significance for a given source.} value $<$ 9), we placed a 95$\%$ confidence upper limit on the photon
flux above 100 MeV, as evaluated with Helene's method (Helene 1983).
A rapid GeV flux and TS value increase was observed beginning  2014 June 6 (31 days after periastron, Figure \ref{mos-1}, red points). The GeV flare is apparent to the eye when plotted in a weekly timescale (Figure \ref{mos-1}b). The detailed daily light curve of the 2014 flare is shown in Figure \ref{mos-1}h, \ref{mos-1}j \& \ref{mos-1}l. A large variability is observed in the 0.1--100 GeV band, with multiple peaks in a daily timescale. To compare the 2010 flare and 2014 flare, we reanalyzed the data from the 2010 periastron passage. The weekly and daily light curves showed consistent results with those reported in Abdo et al. (2011), in which the analysis was carried out with Pass 6 data and P6$\_$V3 IRF\footnote{\url{http://fermi.gsfc.nasa.gov/ssc/data/analysis/documentation/Cicerone/Cicerone_LAT_IRFs/IRF_overview.html}}.

In spite of obvious similarities, the two flares show different time evolution. The light curves of the 2010 (blue) and 2014 (red) periastron passage are plotted together in Figure \ref{mos-1}. It is apparent to the eye that in the weekly light curve, the 2010 flare has a higher peak flux and a rapid decreasing evolution (Figure \ref{mos-1}a). The 2014 flare has a lower peak flux, which then persists rather than falling rapidly. The daily light curves show these same trends. The 2010 flare consists of a large outburst in the beginning followed by several weaker flares. The 2014 flare is composed of several sub-flares with similar flux levels.

We calculated the cross-correlation using \emph{crosscor}\footnote{\url{https://heasarc.gsfc.nasa.gov/xanadu/xronos/help/crosscor.html}} in HEASOFT with the daily flux of the 2010 and 2014 flares  (30 days to 80 days after periastron), which reflects the difference of general trend. We found that  the 2014 flare is $2.4\pm 0.6$ days delayed from the 2010 flare. The error is estimated with 10000 simulations by sampling the daily light curve fluxes within the errors and calculating the cross-correlation for each one.

\begin{figure}
\centering
\includegraphics[width=0.5\textwidth]{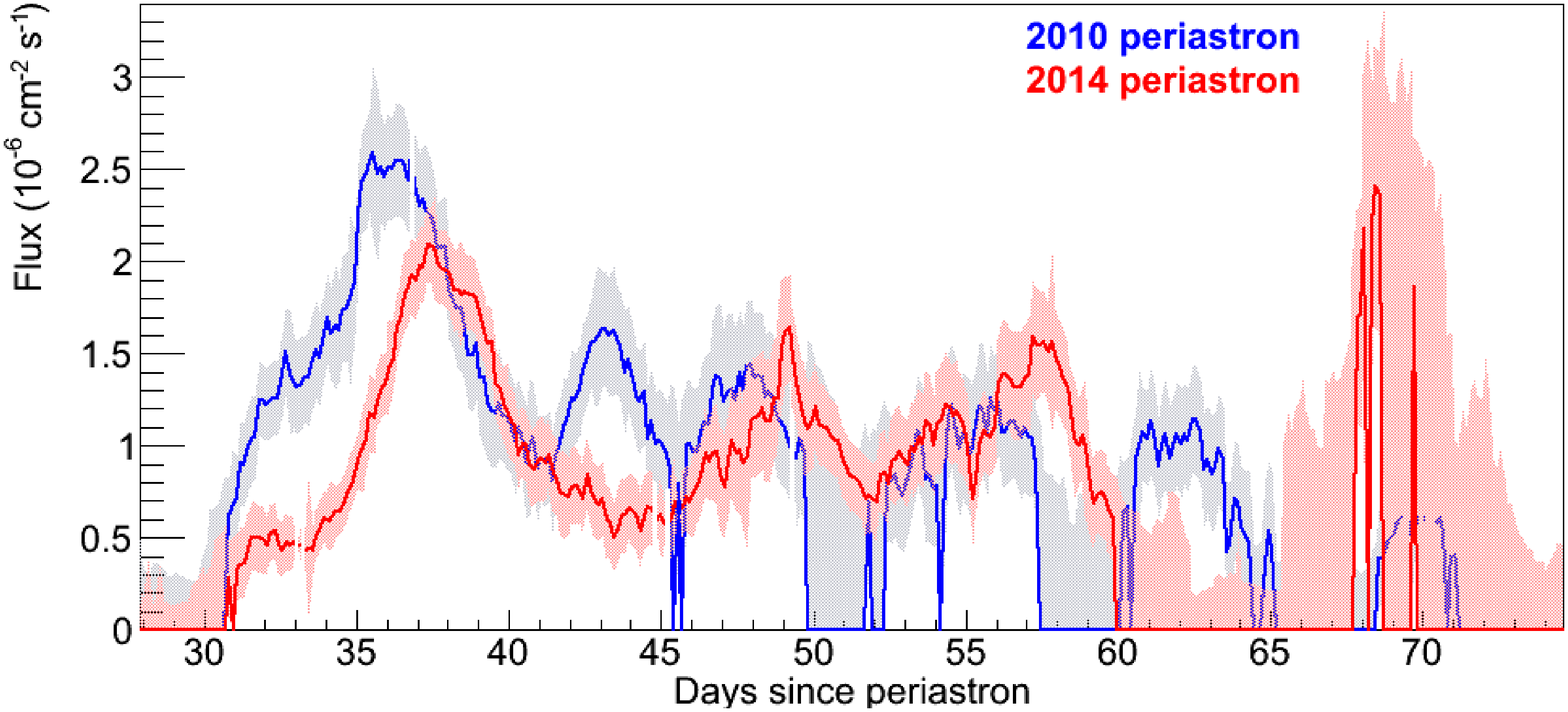}
\includegraphics[width=0.5\textwidth]{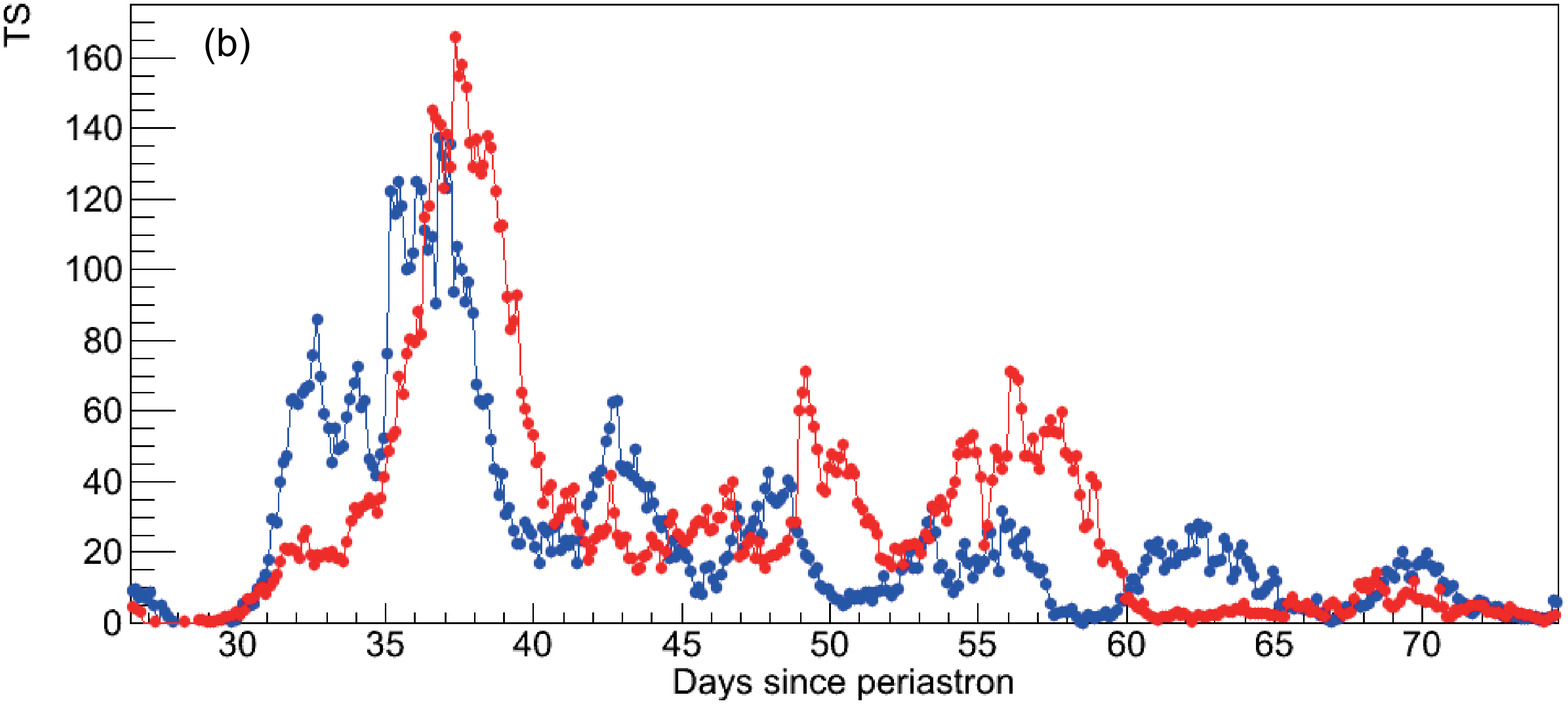}
\caption{Flux (a) and TS (b) smoothed light curves of the 2010 and 2014 flaring activities after periastron (blue and red, respectively) produced with a sliding window. The shaded areas demonstrate the statistical error of the flux. We note that we removed from the plot a few data points where the maximum-likelihood fit did not provide reliable error bars. The smoothed light curve points are represented as null flux when the TS value is below 9, with only shaded areas showing the upper limits at 95$\%$ confidence. See text for more details.}
\label{SmoothedLC}
\end{figure}

In order to better visualize similarities and differences between the 2014 and 2010 flare profiles, their smoothed light curves are plotted in Figure \ref{SmoothedLC}.
The smoothed light curves were produced using a sliding window technique. We chose time windows of 3 days, whose starting times lag the previous one by 3 hours. For each time window, a binned likelihood
analysis was performed. The spectral index of PSR B1259--63 was allowed to vary between the values 2.0 and 3.5.
With the sliding window technique the smoothing is due to the correlation of the points at a time distance smaller than the time window.

Figure \ref{SmoothedLC}a \& \ref{SmoothedLC}b show the flux and the TS smoothed light curves of the 2010 and 2014 periastron passages. In Figure \ref{SmoothedLC}a the shaded areas around the solid lines correspond to the statistical error of the flux calculation. When the TS value is low (TS $<$ 9), the smoothed light curve points are represented as null flux with shaded areas showing the upper limits of 95$\%$ confidence.
In the TS smoothed light curves, the onset of the flaring activity in 2010 and 2014 periastron start approximately at the same time, $\sim 31$ days after periastron. After the onset, the 2010 periastron smoothed flux steeply increases up to its maximum at $\sim 36$ days. However, in 2014 the increase is delayed by forming a short plateau until $\sim 34$ days, after which the flux rises up to the peak (at $\sim 38$ days). The main peak of the 2014 profile is delayed by $\sim 2$ days with respect to the 2010 flare, which is consistent with the result of flux cross-correlation.
A further significant difference is that in 2010 the main peak is followed by a marked second peak (at $\sim 43$ days), which in the 2014 light curve is substituted by a smooth valley. Finally, the 2014 flaring activity ends with a possible peak at $\sim 68$ days, which is consistent with the last high flux bin of the weekly light curve and high upper limits of the daily light curve (see Figure \ref{mos-1}). However, the daily exposure time of this possible peak is $\sim$ 5 times less than the main peak, which leads to low TS values and makes this peak insignificant, preventing further analysis.

We searched for gamma-ray pulsations from \psr\ following the same weighted H-test procedures described in  Abdo et al. (2011) and Hou et al. (2014), using a radio ephemeris internal to the collaboration measured at Parkes Observatory. All LAT observations available excluding periastron passages (from 60 days before periastron until 120 days after periastron) were taken into consideration. Neither a statistically significant detection of a pulsation, nor a  detection of the source was found.  We also searched for gamma-ray pulsation around the 2009 and 2012 apastron passages (from 2, 4 and 6 month intervals centered on apastron). Neither pulsation nor the source itself were significantly detected in these periods. Gamma-ray pulsations in the timeframe around the 2010 and 2014 periastron passages (from 60 days before periastron until 120 days after periastron) were also searched for, but without a detection.

\subsection{Multiwavelength view of both periastra}

\begin{figure*}
\centering
\includegraphics[angle=0, scale=0.65] {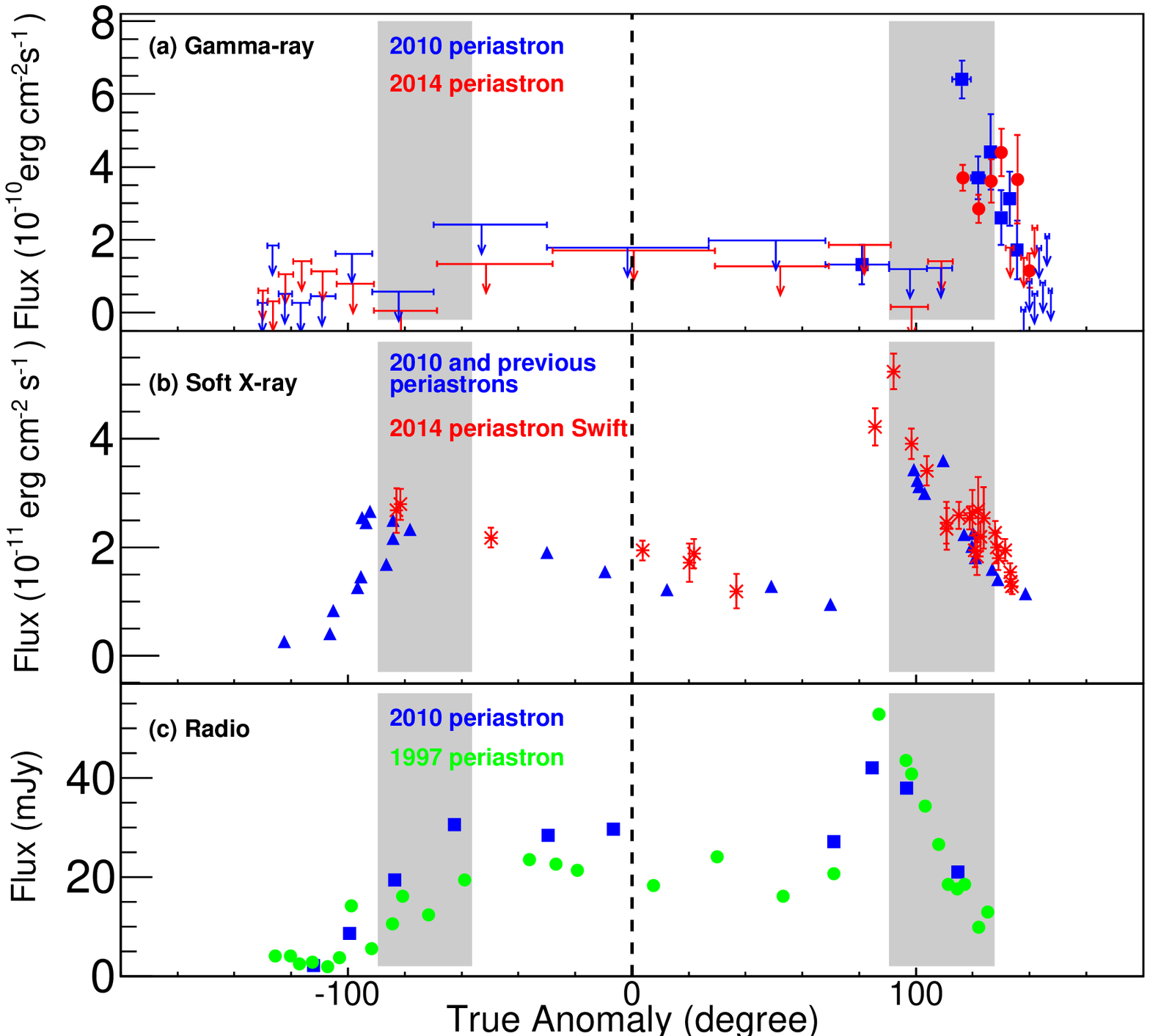}
\caption{Light curves of PSR B1259--63 around periastron. (a) weekly \emph{Fermi}-LAT gamma-ray flux of 2010 (blue) and 2014 (red) periastron passage in energy units. (b) X-ray fluxes from 2014 Swift (red) and previous periastron passages (blue, \emph{XMM-Newton, Chandra, Swift} and \emph{Suzaku } observations during 2004, 2007 and 2010 periastra, Chernyakova et al. 2009. The errors are smaller than the symbols). (c) radio (2.4 GHz) flux densities for the 2010 (blue) and 1997 periastron passages (green, Abdo et al. 2011). X axis is in true anomaly. The dashed black line shows the time of periastron. Shaded area corresponds to the Be circumstellar disk position proposed in Chernyakova et al. (2006).}
\label{mos-5}
\end{figure*}

\psr\ was monitored by \emph{Swift}/XRT during its 2014 periastron. No significant flux increase on timescales shorter than $\sim$4 days was observed in soft X-rays during the GeV flare (Chernyakova et al. 2015). The light curve follows the general trend of previous periastron measurements (Figure \ref{mos-5}b). X-ray measurements of \psr\ during 2014 and previous periastra, and radio measurements during 2010 and 1997 periastra all show a double peaked structure (Figure \ref{mos-5}b \& \ref{mos-5}c), which is expected  when the pulsar crosses the companion's circumstellar disk twice each orbit (Cominsky et al. 1994; Chernyakova et al. 2006, 2009; Abdo et al. 2011). In Figure \ref{mos-5} the position of the Be circumstellar disk is shown with shaded areas as proposed in Chernyakova et al. (2006)

In the 2010 and 2014 first crossings of the Be circumstellar disk, the gamma-ray flux measurements are upper limits because of the low TS value. Although the 2010 and 2014 flares are within the second crossings of the Be circumstellar disk, they are not in line with X-ray and radio activities (Figure \ref{mos-5}a).

\begin{table}
\centering
\caption{Power-law spectrum parameters of PSR B1259--63 during 2014 and 2010 flares}
\label{Table1}
\begin{tabular}{@{}llcc@{}}
\hline
     &   & Photon index &  Flux($>$100 MeV)  \\
     &   & $\Gamma$     &  10$^{-7}$ ph cm$^{-2}$ s$^{-1}$  \\
\hline
             & Average  & 2.93 $\pm$ 0.07 & 9.8 $\pm$ 0.7  \\
2014 flare   & Peak interval  & 2.99 $\pm$ 0.10 & 11.3 $\pm$ 1.0  \\
             & Tail interval  & 2.87 $\pm$ 0.09 & 8.6 $\pm$ 0.9  \\
\hline
             & Average  & 2.90 $\pm$ 0.07 & 11.8 $\pm$ 0.9  \\
2010 flare   & Peak interval  & 2.98 $\pm$ 0.10 & 18.8 $\pm$ 1.5  \\
             & Tail interval  & 2.83 $\pm$ 0.09 & 8.8 $\pm$ 1.1  \\
\hline
\end{tabular}
\end{table}

\subsection{Spectral analysis}

On the basis of the smoothed light curves we defined two time intervals of the flare profiles.
The first interval includes the initial peak (labeled ``Peak" interval). All the rest of the flare constitutes the second interval (labeled ``Tail" interval). The transitions from Peak to Tail are set when TS values in Figure \ref{SmoothedLC} reach the lowest value soon after the peak.
For the 2010 flare, the transition from Peak to Tail occurred 40 days after the periastron while for the 2014 flare this transition occurred 42 days after periastron (Figure \ref{mos-1}g \& h).
The beginnings of both flares occur at 31 days, which is the start of the TS value \& flux rising in Figure \ref{SmoothedLC}. The ends of both flares occur at 71 days, which is the time when the TS values in Figure \ref{SmoothedLC} finally drop below 10.

\begin{figure}
\centering
\includegraphics[angle=0, scale=0.4] {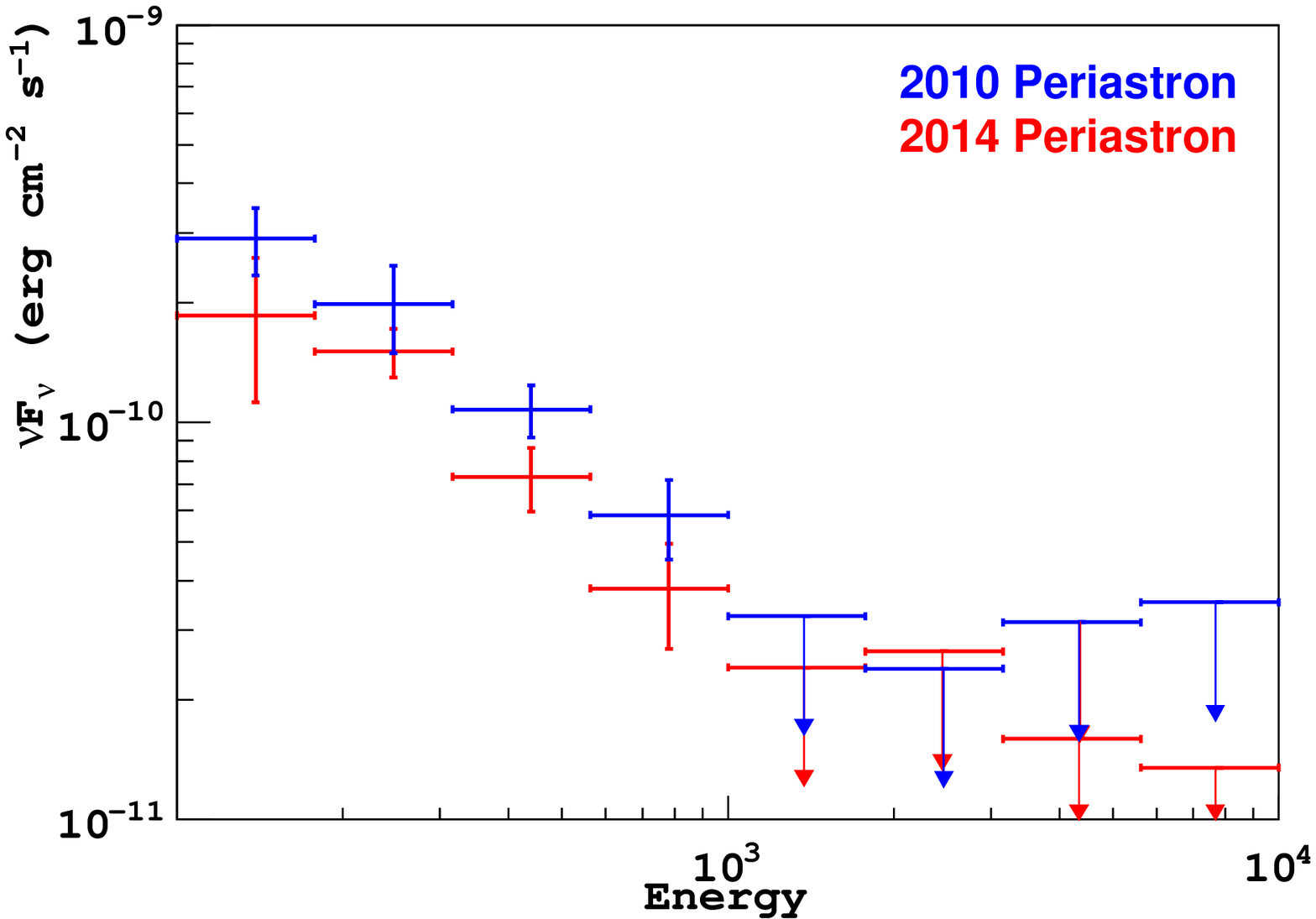}
\includegraphics[angle=0, scale=0.4] {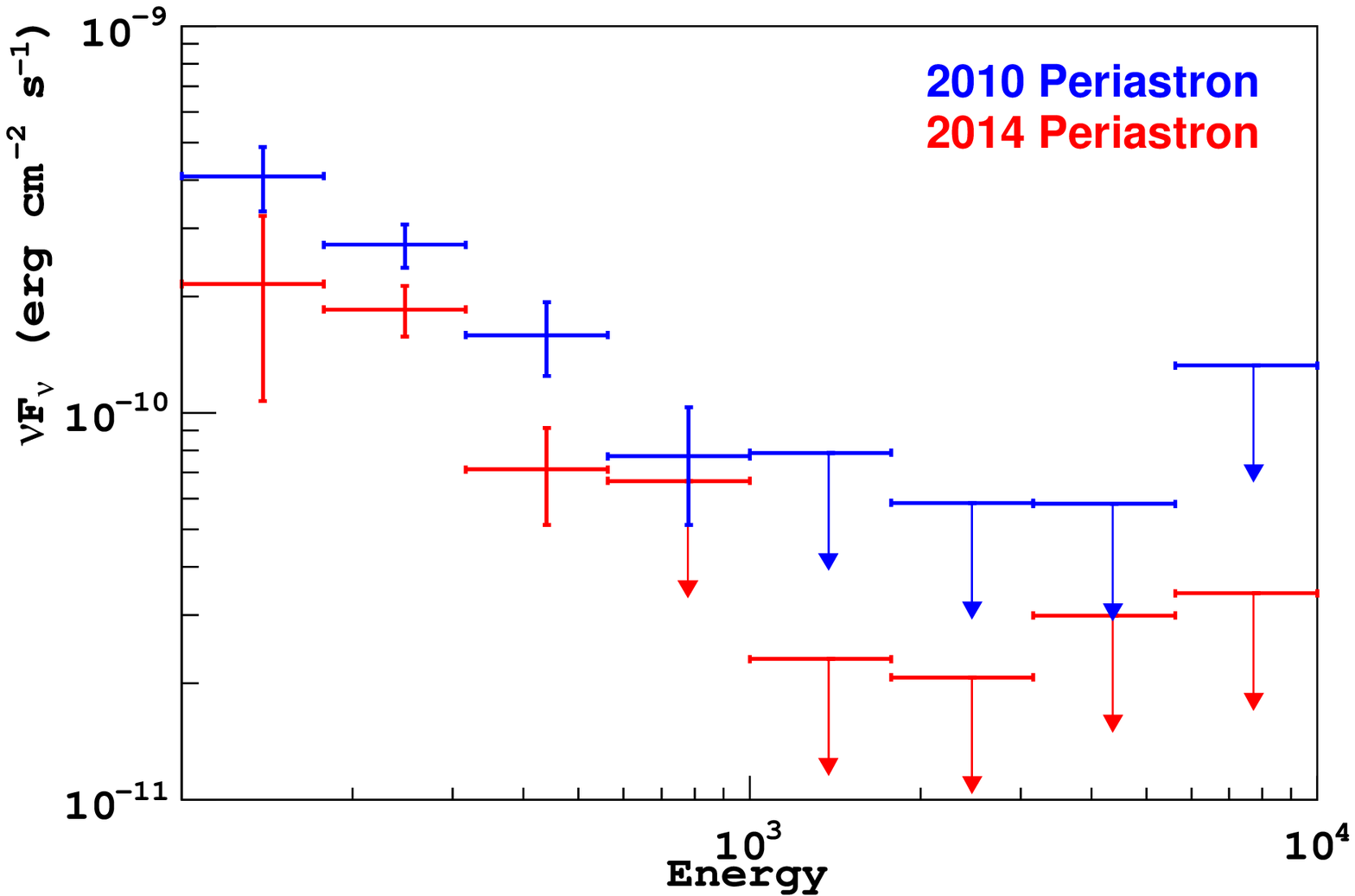}
\includegraphics[angle=0, scale=0.4] {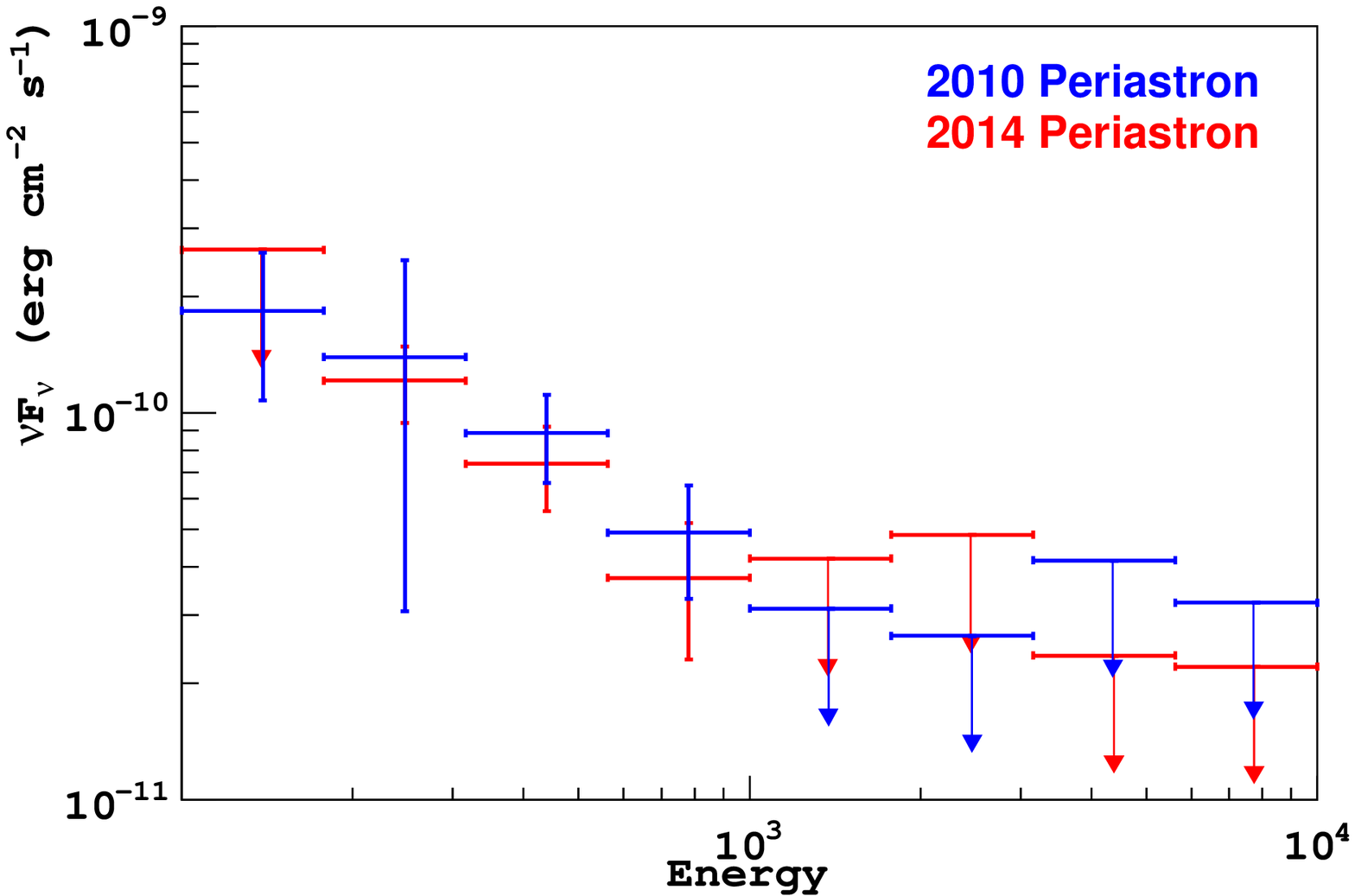}
\caption{\emph{Fermi}-LAT flare spectra of 2010 (blue) and 2014 (red) periastra. (a) Spectra calculated over the whole duration of the flares. (b) Spectra of the Peak intervals of 2010 and 2014 flare. (c) Spectra of the Tail intervals of 2010 and 2014 flare.}
\label{mos-6}
\end{figure}

The average spectra of the whole flares, as well as the spectra of the Peak and Tail intervals were evaluated performing a binned likelihood analysis. The results are shown in Table 1 and Figure \ref{mos-6} . The average spectra during the 2010 and 2014 GeV flare periods are modeled by a single power law and a power law with an exponential cutoff. The spectra are shown in Figure \ref{mos-6}a.  We compare the two models utilizing the likelihood ratio test (Mattox et al. 1996). The $\Delta$ TS between the two models is less than 9, which indicates that the significance of the spectral cutoffs for 2010 and 2014 flares are less than 3$\sigma$. Thus, a power law with an exponential cutoff does not significantly improve the fitting over the simpler power-law model. The fitted parameters of a single power law model are shown in Table 1. The two flares show comparable average flux levels and spectral shape (Table 1; Figure \ref{mos-6}b \& \ref{mos-6}c). No spectral variability is seen above 3$\sigma$ during both 2010 and 2014 flares. The flux of the Peak interval in the 2010 flare is significantly higher than in the 2014 flare (Figure \ref{mos-6}b), while the Tail intervals have similar fluxes (Figure \ref{mos-6}c). In both flares the spectral slope is steeper in the Peak interval than in the Tail, but the variations are not significant.

\section{Discussion}

We have observed the GeV flare during the 2014 periastron passage of \psr\ with \emph{Fermi}-LAT.  These are our main results:

\begin{itemize}

\item We confirm a recurrent flaring behavior in gamma rays, a phenomenon that is associated with the periastron passages.

\item The 2014 GeV flare exhibits a similar average flux level and spectral shape with the 2010 flare.

\item The two GeV flares showed different flux evolution.

\item The 2014 GeV flare peak is about 2 days delayed from the 2010 flare.

\item No GeV pulsations from \psr\ have been detected in any part of its orbit.

\end{itemize}

We have shown that the 2010 and 2014 periastron passages of \psr\ produces some striking similarities in their gamma-ray emission.  At the same time, there
are also certain differences that may be due to inhomogeneities in the shape, density, or extent in the circumstellar disk of the Be star. The two flares were characterized by a high efficiency in the conversion of pulsar spin-down power into gamma-ray emissions.
The 2014 periastron highest day-average flux was $7.1\times10^{-10}$ erg cm$^{-2}$ s$^{-1}$ with a photon index of $\sim$3.1. Though it is lower than the peak day-average flux of the 2010 periastron,  $9.6\times10^{-10}$ erg cm$^{-2}$ s$^{-1}$, it corresponds to an isotropic gamma-ray luminosity of $4.5\times10^{35} $ erg s$^{-1}$, equaling  $\sim 50\%$ of the pulsar spin-down luminosity, $8.3\times10^{35}$ erg s$^{-1}$ (Johnston et al. 1992), without considering possible beaming effects. Gamma-ray conversion efficiency of most gamma-ray emitting pulsars is below $30\%$ (Abdo et al. 2013). Though we have now observed two similar GeV flares from \psr\/, the origin of these events is still unclear. Khangulyan et al. (2012) account for the GeV flare by the upscattering of the circumstellar disk IR photons by the unshocked pulsar wind electrons, which will increase after the pulsar exits the second crossing of the disk. Kong et al. (2012) explain the GeV flare as the Doppler boosted synchrotron emission in the bow-shock tail, which could accommodate the high conversion efficiency of pulsar spin-down power. {Dubus \& Cerutti (2013) adopted upscattering of X-ray photons from the pulsar wind nebula, which would produce correlated X-ray and gamma-ray emission. Such correlation was not observed, which could be due to Doppler boosting or a realistic scattering geometry of the system. The unresolved nature and competing models of GeV flares will motivate future observations dedicated to \psr\/ and hopefully improve our understanding of its unique phenomenology.

\section*{Acknowledgments}
The \textit{Fermi} LAT Collaboration acknowledges generous ongoing support
from a number of agencies and institutes that have supported both the
development and the operation of the LAT as well as scientific data analysis.
These include the National Aeronautics and Space Administration and the
Department of Energy in the United States, the Commissariat \`a l'Energie Atomique
and the Centre National de la Recherche Scientifique / Institut National de Physique
Nucl\'eaire et de Physique des Particules in France, the Agenzia Spaziale Italiana
and the Istituto Nazionale di Fisica Nucleare in Italy, the Ministry of Education,
Culture, Sports, Science and Technology (MEXT), High Energy Accelerator Research
Organization (KEK) and Japan Aerospace Exploration Agency (JAXA) in Japan, and
the K.~A.~Wallenberg Foundation, the Swedish Research Council and the
Swedish National Space Board in Sweden.

Additional support for science analysis during the operations phase is gratefully acknowledged from the Istituto Nazionale di Astrofisica in Italy and the Centre National d'\'Etudes Spatiales in France.

The Parkes radio telescope is part of the Australia Telescope which is funded by the Commonwealth
Government for operation as a National Facility managed by CSIRO. We thank our colleagues for their assistance with the radio timing observations.
J.L. and D.F.T. acknowledge support from the grants AYA2012-39303, SGR 2014-1073 and  support from the National Natural Science Foundation of
China via NSFC-11473027. D.F.T. further acknowledges the Chinese Academy of Sciences visiting professorship program 2013T2J0007.
Work by C.C.C. at NRL is supported in part by NASA DPR S-15633-Y.

\end{document}